\documentclass[twocolumn,floatfix,showpacs,preprintnumbers,superscriptaddress,amsmath,amssymb,prb,citeautoscript]{revtex4}

\usepackage{graphicx}
\usepackage{bm}

\begin{document}


\title{\boldmath Unconventional spin freezing and fluctuations\\ in the frustrated antiferromagnet NiGa$_2$S$_4$}
\author{D.~E. MacLaughlin}
\affiliation{Department of Physics and Astronomy, University of California, Riverside, California 92521}
\author{Y. Nambu}
\affiliation{Institute for Solid State Physics, University of Tokyo, Kashiwa 277-8581, Japan}
\author{S. Nakatsuji}
\affiliation{Institute for Solid State Physics, University of Tokyo, Kashiwa 277-8581, Japan}
\author{R.~H.~Heffner}
\affiliation{Advanced Science Research Center, Japan Atomic Energy
Agency, Tokai 319-1195, Japan}
\affiliation{Los Alamos National Laboratory, Los Alamos, New Mexico 87545}
\author{Lei Shu}
\altaffiliation[Present address: ]{Department of Physics and Institute for Pure and Applied Physical Sciences, University of California, San Diego, La Jolla, California 92093.}
\affiliation{Department of Physics and Astronomy, University of California, Riverside, California 92521}
\author{O. O. Bernal}
\affiliation{Department of Physics and Astronomy, California State University, Los Angeles, California 90032}
\author{K. Ishida}
\affiliation{ Department of Physics, Graduate School of Science, Kyoto University, Kyoto 606-8502, Japan}

\date{October 30, 2008}
 
\begin{abstract}

Muon spin rotation ($\mu$SR) experiments reveal unconventional spin freezing and dynamics in the two-dimensional (2D) triangular lattice antiferromagnet~NiGa$_2$S$_4$. Long-lived disordered Ni-spin freezing (correlation time $\gtrsim 10^{-6}$~s at 2~K) sets in below $T_f = 8.5 \pm 0.5$~K with a mean-field-like temperature dependence. The observed exponential temperature dependence of the muon spin relaxation above $T_f$ is strong evidence for 2D critical spin fluctuations. Slow Ni spin fluctuations coexist with quasistatic magnetism at low temperatures but are rapidly suppressed for $\mathrm{fields} \gtrsim 10$~mT, in marked contrast with the field-independent specific heat. The $\mu$SR and bulk susceptibility data indicate a well-defined 2D phase transition at $T_f$, below which NiGa$_2$S$_4$ is neither a conventional magnet nor a singlet spin liquid.

\end{abstract}

\maketitle

The layered chalcogenide~NiGa$_2$S$_4$ is an example of a quasi-two-dimensional (2D) antiferromagnet (AFM) with a nearly perfect triangular lattice, and is a candidate for a geometrically-frustrated spin liquid \cite{NNTS05}. Its crystal structure consists of triangular Ni planes well separated by GaS polyhedra, which assure the quasi-2D nature of the Ni magnetism. The Ni$^{2+}$ $t^6_{2g}e^2_g$ configuration is that of a $S = 1$ Heisenberg magnet, consistent with the nearly isotopic susceptibility. Neutron scattering experiments \cite{NNTS05,NNOJ07} revealed the development of quasistatic (correlation time~$\tau_c \gtrsim3 \times 10^{-10}$~s) short-range incommensurate Ni spin correlations below ${\sim}$20~K, well below the paramagnetic Curie-Weiss temperature~$|\theta_W| \approx 80$~K, with very weak correlations between Ni planes. The magnetic specific heat~$C_M(T)$ exhibits a broad peak \cite{NNTS05} at $\sim$10~K, and the magnetic susceptibility shows a sharp kink at 8.5~K suggesting a transition at this temperature. Evidence for spin freezing below $\sim$10~K was obtained from $^{69}$Ga NQR experiments \cite{TIKI08} and substantiated by recent muon spin rotation ($\mu$SR) data \cite{YDdRCM08}. 

In the temperature range~0.35--4~K $C_M(T) \propto T^2$, indicative of gapless linearly-dispersive low-lying modes. Magnons in an ungapped 2D AFM would yield this result but would require long-range spin order, in apparent conflict with the neutron scattering results. At all temperatures $C_M(T)$ is independent of applied magnetic field up to 7~T, as also observed in the 2D kagom\'e lattice AFM~SrCr$_{9p}$Ga$_{12-9p}$O$_{19}$ (SCGO) (\mbox{Ref.\ }\onlinecite{RHW00}). This is certainly not expected for magnons or any other simple cooperative spin excitation, and suggests the possibility of a singlet or singlet-like spin liquid. A number of theoretical scenarios have been proposed for NiGa$_2$S$_4$ (\mbox{Refs.\ }\onlinecite{LMP06} and\mbox{\ }\onlinecite{KaYa07}), with magnetic properties that depend sensitively on details of the assumed model as is common in frustrated systems. It is important, therefore, to characterize the properties of this compound as fully as possible. 

This Letter reports a detailed longitudinal-field $\mu$SR (LF-$\mu$SR) (\mbox{Ref.\ }\onlinecite{Brew94}) study of NiGa$_2$S$_4$, using the local nature of the muon spin probe to provide unique information on the static and dynamic behavior of Ni spins at low temperatures. Although preliminary $\mu$SR results \cite{MHNN07} seemed to be inconsistent with spin freezing, the present data and those of \mbox{Ref.\ }\onlinecite{YDdRCM08} give clear evidence for a disordered quasistatic \cite{quasistatic} Ni-spin configuration. We find a lower bound on $\tau_c$ more than three orders of magnitude longer than the neutron scattering value. This spin freezing sets in abruptly and exhibits disordered mean-field-like behavior below a freezing temperature~$T_f \approx 8.5$~K\@. 
A kink is observed in the bulk susceptibility at the same temperature; together these results suggest a well-defined 2D phase transition.
Dynamic muon spin relaxation is spatially inhomogeneous at all temperatures, exhibits an exponential temperature dependence for $T > T_f$ as expected for 2D Heisenberg critical fluctuations, and remains strong down to $\sim$2~K for small applied fields. Our most unexpected finding is a significant suppression of this low-temperature relaxation by fields $\sim 10$~mT, which is difficult to reconcile with the field independence of the specific heat. 

Time-differential LF-$\mu$SR measurements were carried out at the M20 beam line at \mbox{TRIUMF}, Vancouver, Canada, on  a homogeneous and stoichiometric polycrystalline powder and a 
single crystal of NiGa$_2$S$_4$ (\mbox{Ref.\ }\onlinecite{NNTS05}), the latter grown using a chemical vapor transport method. 
Data were taken in a longitudinal field~$\mu_0H_L \ge 2$~mT ($\mathbf{H}_L \parallel [001]$ for the single crystal), to decouple the muon spins from nuclear dipolar fields \cite{HUIN79} and to study the field dependence. Care was taken to minimize the spectrometer ``dead time'' between a muon stop and the earliest detection of the decay positron. This was necessary since the early-time relaxation in NiGa$_2$S$_4$ is very rapid; it was not observed in earlier $\mu$SR experiments at the KEK pulsed muon facility \cite{TIKI08} due to the relatively long dead time inherent in pulsed $\mu$SR\@. 

\begin{figure}[ht]
\includegraphics*[clip=,width=0.45\textwidth]{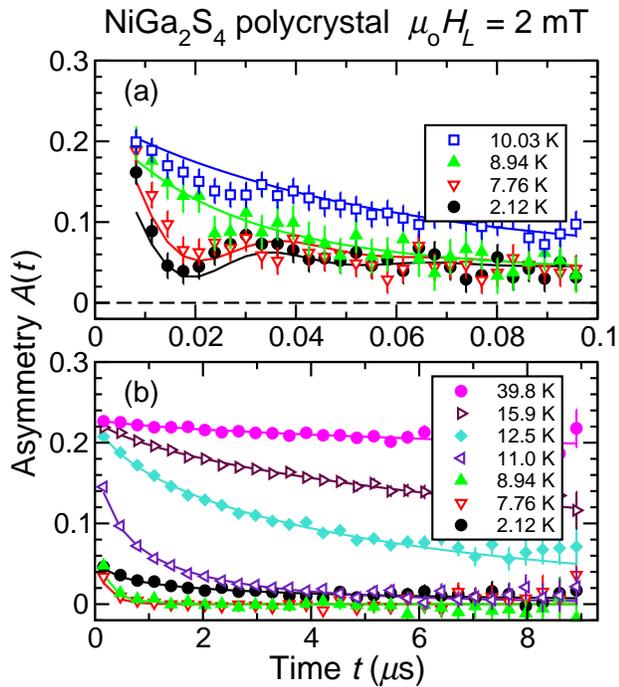}
\caption{(color online) Representative (a)~early- and (b)~late-time LF-$\mu$SR asymmetry data in polycrystalline NiGa$_2$S$_4$, applied longitudinal field~$\mu_0H_L = 2$~mT\@. Curves: fits to Eqs.~(\protect\ref{eq:fiteq}--\ref{eq:powexp}).}
\label{fig:NN86C-2-bp-asy-LF20G}
\end{figure}

Figure~\ref{fig:NN86C-2-bp-asy-LF20G} shows representative LF-$\mu$SR asymmetry decay data \cite{Brew94} in polycrystalline NiGa$_2$S$_4$.
The asymmetry~$A(t)$ is proportional to the muon spin polarization~$P(t)$. In agreement with previous results \cite{YDdRCM08}, the early-time data at low temperatures exhibit strongly damped oscillations, indicative of a quasistatic component~$\langle\mathbf{B}_{\rm loc}\rangle$ of the local field at the muon site. The oscillations are rapidly damped [Fig.~\ref{fig:NN86C-2-bp-asy-LF20G}(a)] due to inhomogeneity in $\langle\mathbf{B}_{\rm loc}\rangle$, leaving a second signal that relaxes slowly [Fig.~\ref{fig:NN86C-2-bp-asy-LF20G}(b)]. This slow signal is due to the muon spin component parallel to $\langle\mathbf{B}_{\rm loc}\rangle$ (Kubo-Toyabe behavior), and its relaxation is 
due to the fluctuating component~$\delta\mathbf{B}_{\rm loc}(t) = \mathbf{B}_{\rm loc}(t) - \langle\mathbf{B}_{\rm loc}\rangle$ (\mbox{Ref.\ }\onlinecite{HUIN79}). 

It should be noted that this two-component Kubo-Toyabe structure arises when the statistical properties of $\mathbf{B}_{\rm loc}$ [inhomogeneity in $\langle\mathbf{B}_{\rm loc}\rangle$, fluctuation statistics associated with $\delta\mathbf{B}_{\rm loc}(t)$] are the same at all muon sites, i.e., when the system is macroscopically homogeneous. Thus the two components should not be associated with separate volume fractions of a multiphase sample, and are not by themselves evidence for multiple phases in NiGa$_2$S$_4$. We discuss below independent evidence for inhomogeneous spin dynamics obtained from the form of the slow-signal relaxation.

The oscillations and Kubo-Toyabe behavior are strong evidence for a quasistatic Ni-spin configuration in NiGa$_2$S$_4$. As $T_f = 8.5 \pm 0.5$~K is approached from below, the oscillation frequency decreases rapidly and the two-component behavior is lost. For $T > T_f$ there is only a single signal component, with a rapidly-increasing rate as $T \rightarrow T_f$ as expected from critical slowing of Ni-spin fluctuations.

For $T < T_f$ the asymmetry data were fit to the form
\begin{equation}
A(t) = A_r P_r(t) + A_s P_s(t) \,, \ P_r(0) = P_s(0) = 1 \,,
\label{eq:fiteq}
\end{equation}
where the the first and second terms represent the rapid and slow components, respectively (``$T < T_f$ fits''). The total asymmetry~$A = A_r + A_s$ at $t = 0$ was assumed independent of temperature and applied field. A damped Bessel function 
\begin{equation}
P_r(t) = \exp(-\lambda_r t) J_0(\omega_\mu t) \,,
\end{equation}
indicative of an incommensurate spin structure \cite{LKLS93}, was found to fit the early-time data better than other candidates such as a damped sinusoid~$\exp(-\lambda_r t) \cos(\omega_\mu t)$ or simple Kubo-Toyabe functions \cite{HUIN79,UYHS85}. The spectrometer dead time of $\sim$8~ns makes the choice of fit function uncertain, but the fit value of the oscillation frequency~$\omega_\mu$ does not depend strongly on this choice. For the Bessel-function fit $\omega_\mu/2\pi = 34 \pm 2$~MHz, $\langle B_{\rm loc}\rangle = \omega_\mu/\gamma_\mu = 251 \pm 15$~mT, where $\gamma_\mu$ is the muon gyromagnetic ratio. This value is qualitatively consistent with Ni dipolar fields at candidate muon stopping sites in the GaS layers. The ratio~$\lambda_r/\omega_\mu \approx 0.2$ is a measure of the relative spread in $\langle B_{\rm loc}\rangle$ due to disorder.

The late-time asymmetry data below $T_f$ and the entire asymmetry function above $T_f$ ($A_r = 0$, ``$T > T_f$ fits'') could be well fit with the ``stretched exponential'' relaxation function
\begin{equation}
P(t)\ [= P_s(t)\ \mbox{for}\ T < T_f] = \exp[-(\Lambda t)^K] \,, \ K < 1 \,,
\label{eq:powexp}
\end{equation}
often used \cite{KMCL96} to model an inhomogeneous distribution~$p(W)$ of exponential rates~$W$.\cite{distrelax} The relaxation time~$1/\Lambda$ gives the time scale of the relaxation and the power~$K$ controls its shape: broader distributions of relaxation rates are modeled by smaller values of $K$\@. For all fits $K$ was found to be significantly smaller than 1, indicating that the spin fluctuations are inhomogeneous over the entire temperature range \cite{nodecoupling}. At low temperatures and low fields the amplitude~$A_s$ of the slowly-relaxing component is approximately $A/3$, as expected from quasistatic LF-$\mu$SR relaxation in a powder sample \cite{HUIN79}.

\begin{figure}[ht]
\includegraphics*[clip=,width=0.45\textwidth]{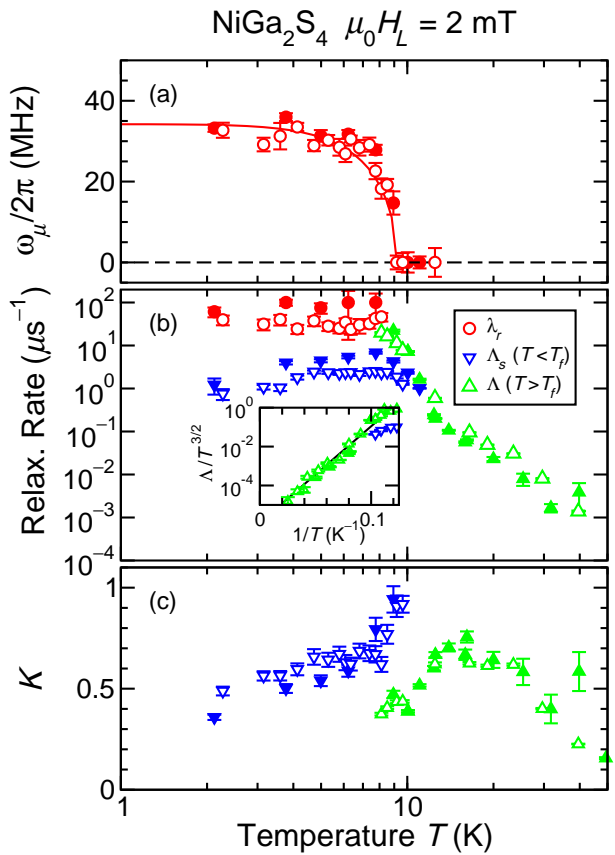}
\caption{(color online) Temperature dependencies of fit parameters in NiGa$_2$S$_4$, $\mu_0H_L = 2$~mT\@. Filled symbols: polycrystalline sample. Open symbols: single-crystal sample, $\mathbf{H}_L \parallel [001]$. Red symbols: (a): oscillation frequency~$\omega_\mu/2\pi$. (b): static relaxation rate~$\lambda_r$. Blue symbols: $T < T_f$ fits (see text); (b)~dynamic rate~$\Lambda_s$, (c)~stretching power~$K$\@. Green symbols: $T > T_f$ fits; (b)~$\Lambda$, (c)~$K$\@. Inset to (b): $T > T_f$, $\Lambda/T^{3/2}$ vs $1/T$\@. Line: $\Lambda/T^{3/2} \propto \exp(T^*/T)$ (\mbox{Ref.\ }\onlinecite{CHN89}), $T^* = 114(7)$~K. \label{fig:nigas-bre-LF20G}}
\end{figure}

Figure~\ref{fig:nigas-bre-LF20G} gives the temperature dependencies of the parameters from fits to data from both single-crystal and polycrystalline samples. 
For the two samples $\omega_\mu(T)$ and $K$ reproduce well. With increasing temperature $\omega_\mu$ decreases and vanishes at $T_f = 8.5 \pm 0.5$~K, in agreement with the kink in the bulk susceptibility \cite{NNTS05}. A mean-field temperature dependence [Brillouin function for $S = 1$, curve in Fig.~\ref{fig:nigas-bre-LF20G}(a)] describes the data to within experimental uncertainty. 

The dynamic relaxation rates in the polycrystalline sample are higher below $T_f$ and lower above $T_f$ than in the single crystal but follow the same trends [Fig.~\ref{fig:nigas-bre-LF20G}(b)]. As $T$ approaches $T_f$ from above $\Lambda$ increases strongly, suggestive of critical slowing down of paramagnetic-state spin fluctuations. The data can be fit to $\Lambda/T^{3/2} \propto \exp(T^*/T)$ [inset of Fig.~\ref{fig:nigas-bre-LF20G}(b)] with $T^* = 114(7)$~K\@. This form is expected for critical fluctuations of Heisenberg spins on a 2D lattice \cite{CHN89} with $T^* = 2\pi\rho_s$, where for $S = 1$ the spin-wave stiffness $\rho_s \approx J$\@. From $\theta_W = zJS(S+1)/3$ with $z = 6$ Ni nearest neighbors, we estimate $T^* = 2\pi\rho_s \approx 125$~K, in good agreement with the measured value. The differences in $\Lambda$ and $K$ for $T < T_f$ and $T > T_f$ fits near $T_f$ [Figs.~\ref{fig:nigas-bre-LF20G}(b) and (c)] probably reflect a distribution of freezing temperatures.

With decreasing temperature below $T_f$ $\Lambda_s$ decreases slightly, but for both samples remains $\gtrsim 1~\mu\mathrm{s}^{-1}$ down to $2.1~\mathrm{K} \approx T_f/4$. This strong relaxation is consistent with the loss of $^{69}$Ga NQR signal in the same temperature range \cite{TIKI08}. Somewhat similar relaxation behavior has been observed in the triangular AFM~NaCrO$_2$ (\mbox{Ref.\ }\onlinecite{OMBU06}) with, however, no indication of critical slowing down above $T_f \approx 40$~K\@.

\begin{figure}[t]
\includegraphics*[clip=,width=0.45\textwidth]{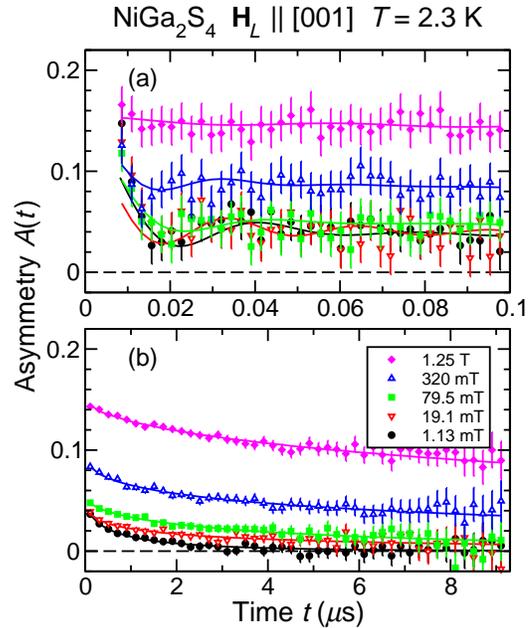}
\caption{(color online) Representative (a)~early- and (b)~late-time LF-$\mu$SR asymmetry plots in single-crystal NiGa$_2$S$_4$ at $T = 2.3$~K and applied longitudinal fields~$\mathbf{H}_L \parallel [001]$.}
\label{fig:na308S2-bp-asy-LF2K}
\end{figure}

\begin{figure}[ht]
\includegraphics*[clip=,width=0.45\textwidth]{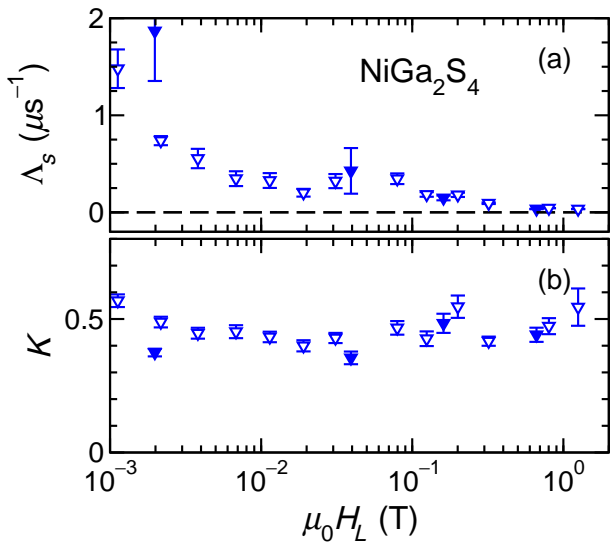}
\caption{(color online) Dependencies of (a)~slow relaxation rate~$\Lambda_s$ and (b)~stretching power~$K$ [Eq.~(\protect\ref{eq:powexp})] on applied longitudinal field~$H_L$ in NiGa$_2$S$_4$. Filled symbols: polycrystalline sample, $T = 2.1$~K\@. Open symbols: single-crystal sample, $T = 2.3$~K, $H_L = \parallel [001]$.}
\label{fig:nigas-bre-LF2K}
\end{figure}

LF-$\mu$SR asymmetry data from the single-crystal sample at 2.3~K are shown in Fig.~\ref{fig:na308S2-bp-asy-LF2K} for $\mu_0H_L$ in the range~1.13~mT--1.25~T\@.
The rapid early-time relaxation is decoupled (the late-time amplitude increases) for $\mu_0H_L \gtrsim \langle B_{\rm loc}\rangle$. This decoupling is independent evidence for a distribution of quasistatic fields \cite{HUIN79}, and is in contrast to the ``undecouplable Gaussian'' relaxation observed in $\mu$SR experiments on SCGO (\mbox{Ref.\ }\onlinecite{UKKL94}). Figure~\ref{fig:nigas-bre-LF2K} shows the field dependencies of $\Lambda_s$ and $K$ at 2.1~K (polycrystal) and 2.3~K (single crystal). 
The stretching power~$K \approx 0.4$ is more or less independent of field. It can be seen that $\Lambda_s$ is strongly suppressed by applied field $\mu_0H_L \gtrsim 10~\mbox{mT} \ll \langle B_{\rm loc}\rangle$. This suppression is not due to an effect of $H_L$ on the muon Zeeman splitting \cite{KMCL96}, because the resultant muon field~$\mu_0\mathbf{H}_L + \langle \mathbf{B}_{\rm loc}\rangle$ is still approximately $\langle \mathbf{B}_{\rm loc}\rangle$. Thus $H_L$ must affect the Ni spin fluctuation spectrum directly. This contrasts sharply with the field-independent specific heat, which is unaffected to within a few percent by a field of 7~T\@. 

We conclude that NiGa$_2$S$_4$, which should be a good example of a 2D triangular Heisenberg $S = 1$ magnet, is neither a ``conventional'' AFM, for which spin freezing would be accompanied by a decrease of the muon relaxation at low temperatures, nor a singlet spin liquid, for which there would be no spin freezing. The LF-$\mu$SR data are consistent with a well-defined 2D phase transition at $T_f \approx 8.5$~K leading to a disordered AFM ground state, as found from previous $^{69}$Ga NQR and $\mu$SR studies. \cite{TIKI08,YDdRCM08} This is in contrast to SCGO, which exhibits a field-independent specific heat \cite{RHW00} but no spin freezing \cite{UKKL94}. The low-temperature relaxation rate~$\Lambda_s$ gives a lower bound on the frozen-spin correlation time~$\tau_c$ of $1/\Lambda_s \sim 10^{-6}$~s at 2~K, more than three orders of magnitude longer than the lower bound from neutron scattering and seven orders of magnitude longer than the inverse exchange frequency. Thus the Ni spins are very viscous indeed, if not completely frozen. The transition at $T_f$ is unlikely to be due either to 3D coupling or to Ni-spin anisotropy, both of which are weak;\cite{NNTS05} a topological transition associated with vortex binding \cite{KaYa07} is one candidate mechanism. NiGa$_2$S$_4$ is chemically and structurally well ordered, raising the question of why the $\mu$SR and $^{69}$Ga NQR relaxation \cite{TIKI08} is spatially inhomogeneous.

The persistent muon spin relaxation at low temperatures [Fig.~\ref{fig:nigas-bre-LF20G}(b)] is evidence for significant Ni spin fluctuation noise power at low frequencies. This is often observed in frustrated magnets without spin freezing \cite{MOBB07} but less frequently in spin-frozen states \cite{BMBO06,OMBU06,YDdRGM05}. It indicates a high density of spin excitations that coexists with spin freezing at low temperatures \cite{NiGaSrelax}. A mean-field-like order parameter is also rare in 2D frustrated antiferromagnets \cite{KKLL96}, and to our knowledge has never been seen together with persistent low-temperature dynamics. 

Easily the most remarkable result of this work is the contrast between the suppression of the low-temperature muon spin relaxation by $H_L \gtrsim 10$~mT [Fig.~\ref{fig:nigas-bre-LF2K}(a)] and the negligible magnetic field dependence of the specific heat up to fields more than two orders of magnitude greater \cite{NNTS05}. There is no fundamental paradox here, since the entropy of a spin system is not explicitly dependent on the time scale of the fluctuations. But it is hard to understand how strong muon spin relaxation, usually associated with most or all of the spin excitation degrees of freedom of a system, could change so drastically with field without the slightest signature in the entropy. Reconciling the LF-$\mu$SR and specific-heat results is clearly an important task for theory. 

\begin{acknowledgments}
We are grateful for technical assistance from the \mbox{TRIUMF} Centre for Molecular and Materials Science, where these experiments were carried out. We wish to thank H. Kawamura, Y. Maeno, and R.~R.~P. Singh for useful discussions, and K. Onuma for help with the experiments. This work was supported by the U.S. NSF, Grants~0422674 (Riverside) and 0604015 (Los Angeles), by Grants-in-Aid for Scientific Research from JSPS, and by a Grant-in-Aid for Scientific Research on Priority Areas (19052003) (Tokyo).
\end{acknowledgments}



\begin{thebibliography}{24}
\bibitem[{qua()}]{quasistatic}
\bibinfo{note}{A muon local-field component~$\langle \mathbf{B}_{\rm
  loc}\rangle$ is quasistatic if it varies slowly compared to the muon spin
  precession period~$2\pi/\gamma_\mu\langle B_{\rm
  loc}\rangle$ (Ref.\mbox{~}\protect\onlinecite{HUIN79}). We include the static limit in our use of
  this term. The dynamic muon spin relaxation time in zero or low applied field
  is then a lower bound on the correlation time of $\langle \mathbf{B}_{\rm
  loc}\rangle$.}
\bibitem[{nod()}]{nodecoupling}
\bibinfo{note}{Reference\mbox{~}\protect\onlinecite{YDdRCM08} reported exponential
  rather than stretched-exponential muon relaxation in zero field above
  $\sim$10~K\@. This difference may be due to nuclear dipolar fields, which
  contribute in zero field (cf.\ Ref.\mbox{~}\protect\onlinecite{TIKI08}) but are decoupled
  in our LF-$\mu$SR experiments.}
\bibitem[{NiG()}]{NiGaSrelax}
\bibinfo{note}{Neutron scattering data~\protect\cite{NNTS05,NNOJ07} indicate a
  quasistatic Ni moment of ${\sim}1.5~\mu_B$, 75\% of the Ni$^{2+}$ $S = 1$
  Hund's-rule value. The remaining component is fluctuating and is presumably
  responsible for the $\mu^+$ relaxation.}
\end{thebibliography}


\end{document}